\documentclass{PoS}

\title{Status of the Daya Bay Reactor Neutrino Oscillation Experiment}

\ShortTitle{Status of the Daya Bay Reactor Neutrino Oscillation Experiment}

\author{\speaker{Cheng-Ju Lin}\\%
       Lawrence Berkeley National Laboratory\\
       E-mail: \email{cjslin@lbl.gov}}

\author{On Behalf of the Daya Bay Collaboration\\}
\abstract{
The last unknown neutrino mixing angle $\theta_{13}$ is one of the fundamental parameters of nature; it is also a crucial parameter for determining the sensitivity of future long-baseline experiments aimed to study CP violation in the neutrino sector.  Daya Bay is a reactor neutrino oscillation experiment designed to achieve a sensitivity on the value of $sin^2(2\theta_{13})$ to better than 0.01 at 90\% CL. The experiment consists of multiple identical detectors placed underground at different baselines to minimize systematic errors and suppress cosmogenic backgrounds.  With the baseline design, the expected anti-neutrino signal at the far site is about 360 events per day and at each of the near sites is about 1500 events per day.  An overview and current status of the experiment will be presented.}

\FullConference{35th International Conference of High Energy Physics\\
                July 22-28, 2010\\
                Paris, France}

\begin{document}

%\section{Introduction}
%Recent neutrino oscillation experiments have provided compelling evidences that neutrinos have non-zero mass.  The Pontecorvo-Maki-Nakagawa-Sakata (PMNS) matrix, which describes the mixing between the neutrino flavor and mass eigenstates, can be characterized by three mixing angles ($\theta_{12}, \theta_{23}, \theta_{13}$), and a Charge-Parity (CP) phase $\delta_{CP}$\footnote{If neutrinos are Majorana particles, there are two additional Majorana phases that are needed in the PMNS matrix.}.  The mixing angles $\theta_{12}$ and  $\theta_{23}$ have been measured by a combination of solar, atmospheric and reactor neutrino experiments.  The last mixing angle $\theta_{13}$ is still unknown and only experimental upper bounds exist.  The mixing angle $\theta_{13}$ is an important fundamental parameter of nature.  Furthermore, it is also a gateway to study CP violation in the neutrino sector.  The goal of the Daya Bay experiment is to pin down the last unknown mixing angle with a sensitivity better than 0.01 in sin$^2(2\theta_{13})$ at the 90\% C.L. by measuring the rates and the energy spectra of the anti-neutrinos from the nuclear reactors at different baselines.  

%\section{Experimental Design and Status}
\section*{}
\begin{figure}[htb]
\begin{center}
\includegraphics[width=0.5\textwidth]{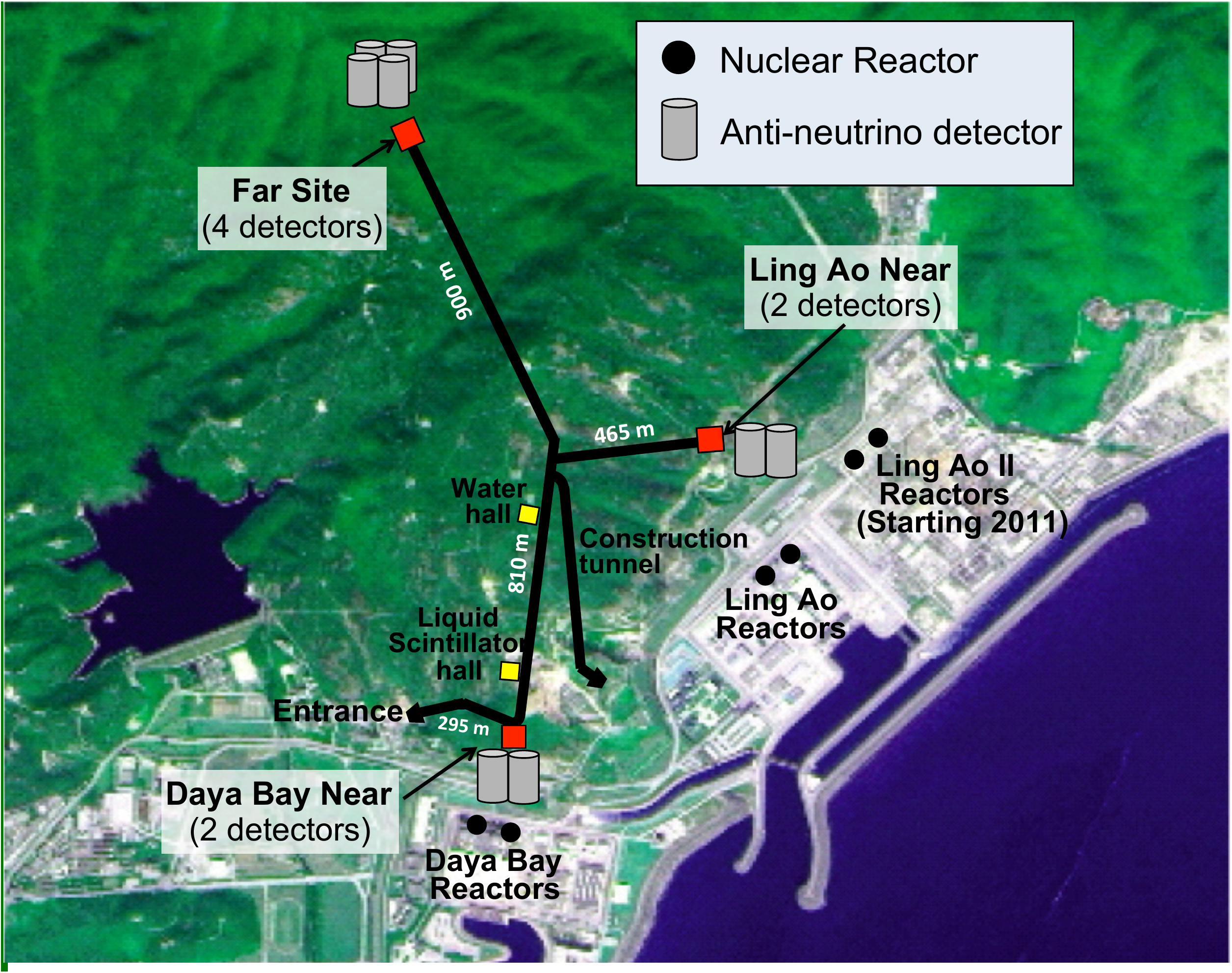}
\caption{Layout of the experimental sites.} \label{fig1}
\end{center}
\end{figure}

The Daya Bay reactor neutrino experiment~\cite{Dayabay} is under construction at the Daya Bay nuclear power plant near Shenzhen, China.  It is an international collaboration consisting of about 240 members.  The goal of the experiment is to pin down the last unknown neutrino mixing angle $\theta_{13}$ with a sensitivity better than 0.01 in sin$^2(2\theta_{13})$ at the 90\% C.L. by measuring the rates and the energy spectra of the anti-neutrinos from the nuclear reactors at different baselines.  Currently the nuclear power complex has two pairs of reactor cores (Daya Bay and LingAo sites) with a combined thermal output of 11.6 GW$_{th}$.  Another pair of reactor cores is expected to come online at LingAo-II site in early 2011, which will increase the total reactor power to 17.4 GW$_{th}$.   Figure~\ref{fig1} shows the layout of the experimental site.  In addition to having an intense anti-neutrino source nearby, the experimental site is situated in a mountainous region that provides natural overburden to shield the underground anti-neutrino detectors from cosmogenic muon backgrounds.  There are two experimental halls near the reactor cores (Daya Bay and Ling Ao sites) for measuring the neutrino flux and one far hall located near the oscillation maximum.  The three underground experimental halls are connected by horizontal tunnels.  The experiment is constructing 8 identical anti-neutrino detectors; two detectors will be placed at each of the near sites, and the remaining four at the far site.
The underground civil construction is progressing well.  The three sites will be commissioned in sequence with the Daya Bay near site ready for physics data taking in the middle of 2011, the Ling Ao site in late 2011, and the far hall in late 2012.  

The Daya Bay anti-neutrino detector (AD) is designed to identify $\bar{\nu}_e$ events via the inverse beta reaction $\bar{\nu}_e + p \rightarrow e^+ + n$.  The detector, as shown in Figure~\ref{fig2}, consists of two cylindrical acrylic vessels, with the smaller 3-m diameter vessel nested inside the larger 4-m diameter vessel.  The inner acrylic vessel (IAV) is filled with 20 tons of Gadolinium doped liquid scintillator (Gd-LS).   The presence of Gd in the IAV increases the neutron capture cross-section in that region.   The experiment is designed to efficiently identify  $\bar{\nu}_e$ interactions in the IAV.  The region between the inner and outer acrylic vessel (OAV) is filled with 20 tons of un-doped liquid scintillator (linear alkylbenzene) to capture $\gamma$ rays escaped from the target region.  Lastly, the acrylic vessels are secured inside a 5-m diameter stainless steel vessel (SSV) with 192 8-inch photomultiplier tubes (PMT) mounted on the inner cylindrical wall.  The area between the OAV and the SSV is filled with mineral oil to suppress natural radioactivity from the PMT glass and other sources.  To improve light collection efficiency, there is an optical reflector mounted on the top and one at the bottom outer-surface of the OAV.  Each AD is equipped with three independent calibration systems mounted on the top of the SSV.  The calibration system is able to deploy LED and radioactive sources inside the detector to monitor and calibrate the performance of the AD.  At the time this paper is written, two ADs have been fully assembled and are stored in the surface assembly building.  The assembly work for the next pair of ADs is in progress.
\begin{figure}
\begin{minipage}[b]{0.5\linewidth}
\centering
\includegraphics[width=2.5in]{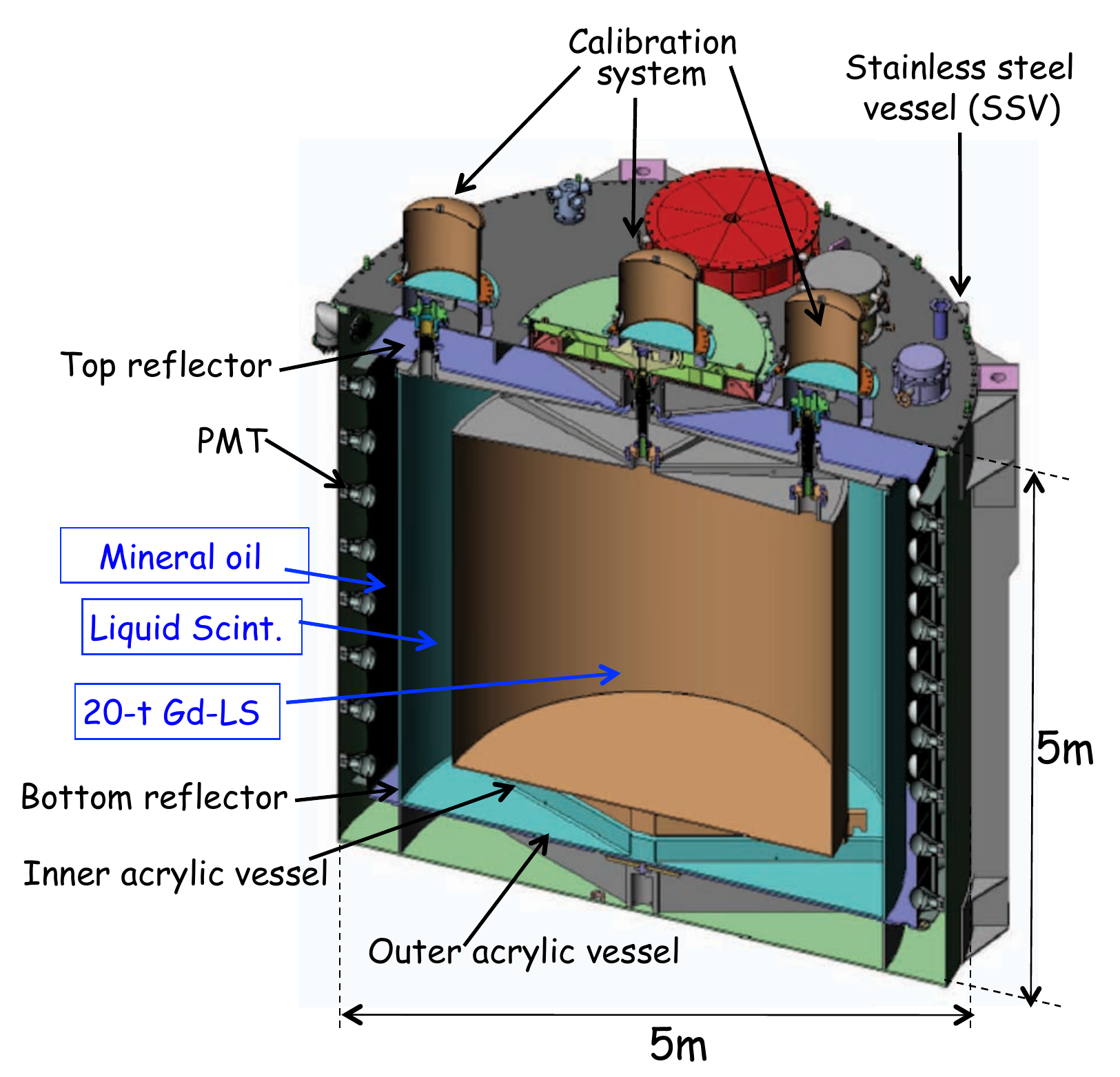}
\caption{Anti-neutrino detector.} \label{fig2}
\end{minipage}
\begin{minipage}[b]{0.5\linewidth}
\centering
\includegraphics[width=2.3in,height=2.2in]{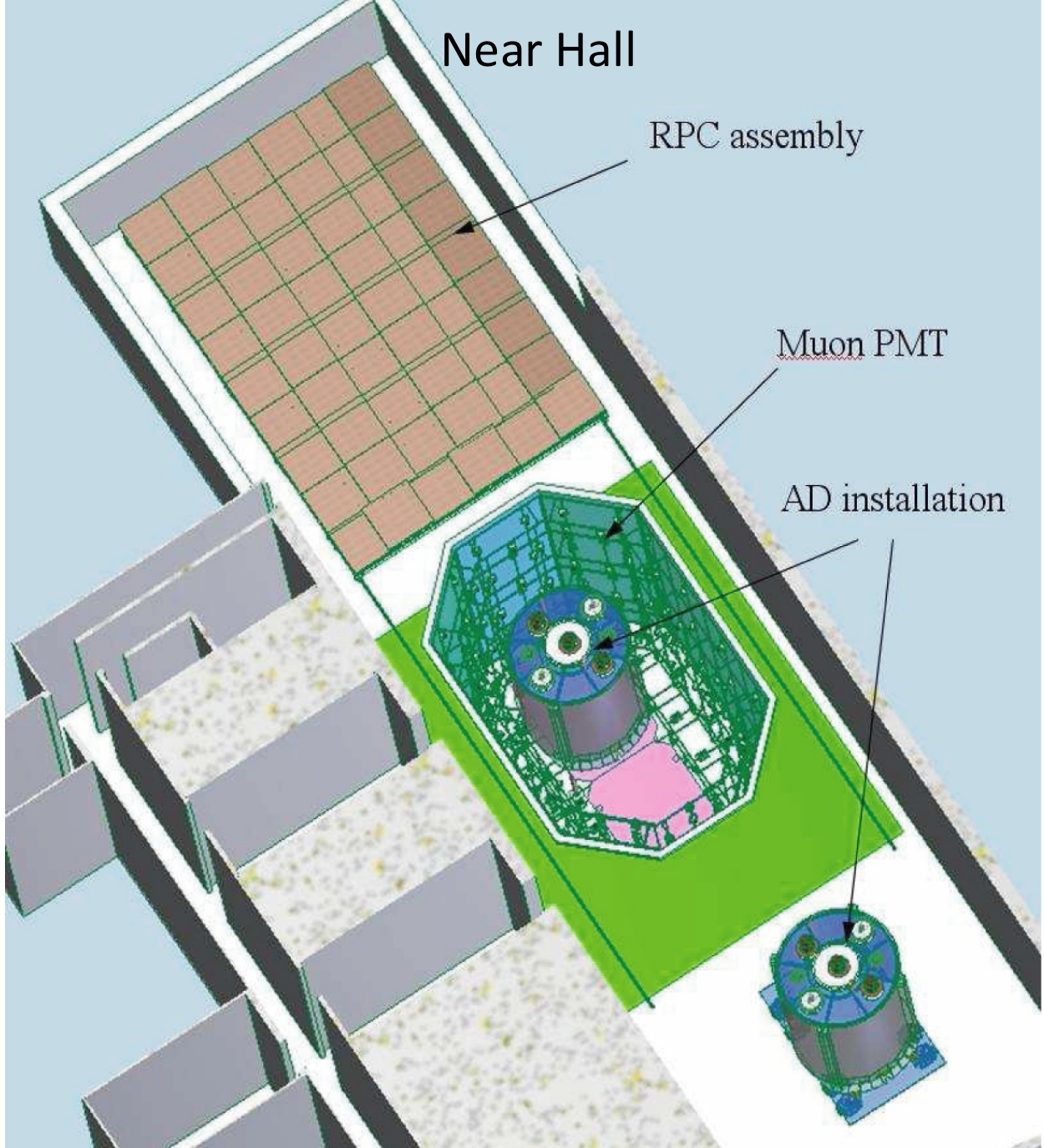}
\caption{Daya Bay near hall.} \label{fig3}
\end{minipage}
\end{figure}

During physics runs, the ADs are immersed in water pool in the underground halls to shield from cosmogenic muon backgrounds and natural radioactivity of the rock.  Figure~\ref{fig3} shows the experimental layout of the near hall.   The design of the far hall is similar to the near halls, with larger dimensions to accommodate the four ADs.  The walls of the water pool are instrumented with PMTs to identify the Cherenkov light from muons traversing the pool.  The water muon system is partitioned into two optically isolated inner and outer zones for redundancy.  There is also a four-layer resistive plate chamber system (RPC) that can be rolled on top of the water pool during data taking to detect cosmic muons.  The estimated combined water Cherenkov and RPC muon detection efficiency is expected to be (99.50$\pm$0.25)\%.

The experiment requires about 200 tons of 0.1\% Gadolinium-doped liquid scintillator.  The primary liquid scintillator solvent is Linear Alkylbenzene.  Linear Alkylbenze has good light yield, high flash point (130$^\circ$ C), and is readily available commercially from manufacturers of biodegradable detergent.  The production of the Gd-LS is taking place in the underground LS hall and is expected to finish by January of 2011.  The stability of the Gd-LS has been monitored in the prototype AD since 2007; no degradation in optical properties has been observed.

%The frontend readout electronics consists of custom-built waveform sampling Analog-Digital-Converter (ADC) for measuring the charge of the PMT hits and  Time-to-Digital-Converter (TDC) for measuring the arrival time of the individual PMT hits.  The 12-bit ADC has two dynamic ranges (high and low gain) to cover the expected signal from neutrino interactions to cosmic muon showers in the AD.  The TDC is integrated in the same electronic module and has a timing resolution of about one nanosecond.  The complete readout system for one AD has been assembled and tested at Daya Bay.  

Recently, we have successfully completed the "dry run" of the first two ADs in the surface assembly building.  The purpose of the dry run is to fully test the individual components of the AD and also to integrate the online and offline infrastructure to simulate physics data taking.  A complete DAQ readout system for one AD was assembled and tested onsite.  During the dry run, data were taken nearly continuously onsite and distributed automatically to offsite analysis centers.  The dry run data have been used to certify that all PMTs for the two ADs are fully functional.  The data are being used to calibrate the performance of the PMTs and the electronics modules (e.g. noise, pedestal, PMT gains, TDC offsets, trigger efficiency, detector uniformity, etc.).

In summary, the construction of the Daya Bay reactor neutrino experiment is well underway with the expected completion date near the end of 2012.  A summary of the uncertainties and sensitivity of the experiment are shown in Table~\ref{tab1} and Figure~\ref{fig4}, respectively.  With three sites operational, the sensitivity on $\theta_{13}$ will reach interesting regions in a very short time and surpass all other experiments in operation in less than a year of data taking.
\begin{figure}
\begin{minipage}[hbt]{0.5\linewidth}
\centering
\includegraphics[width=0.8\textwidth]{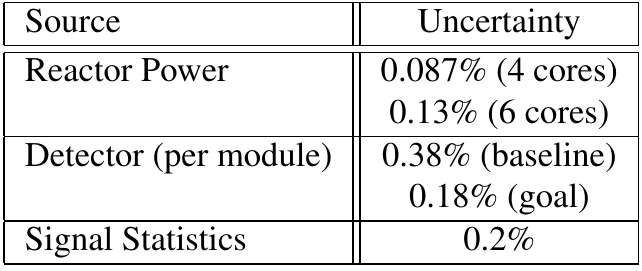}
\setcounter{figure}{0}
\renewcommand{\figurename}{Table}
\caption{Summary of uncertainties on sin$^2(2\theta_{13})$.}\label{tab1}
\end{minipage}
\begin{minipage}[hbt]{0.5\linewidth}
\centering
\setcounter{figure}{3}
\includegraphics[width=2.5in,height=2.05in]{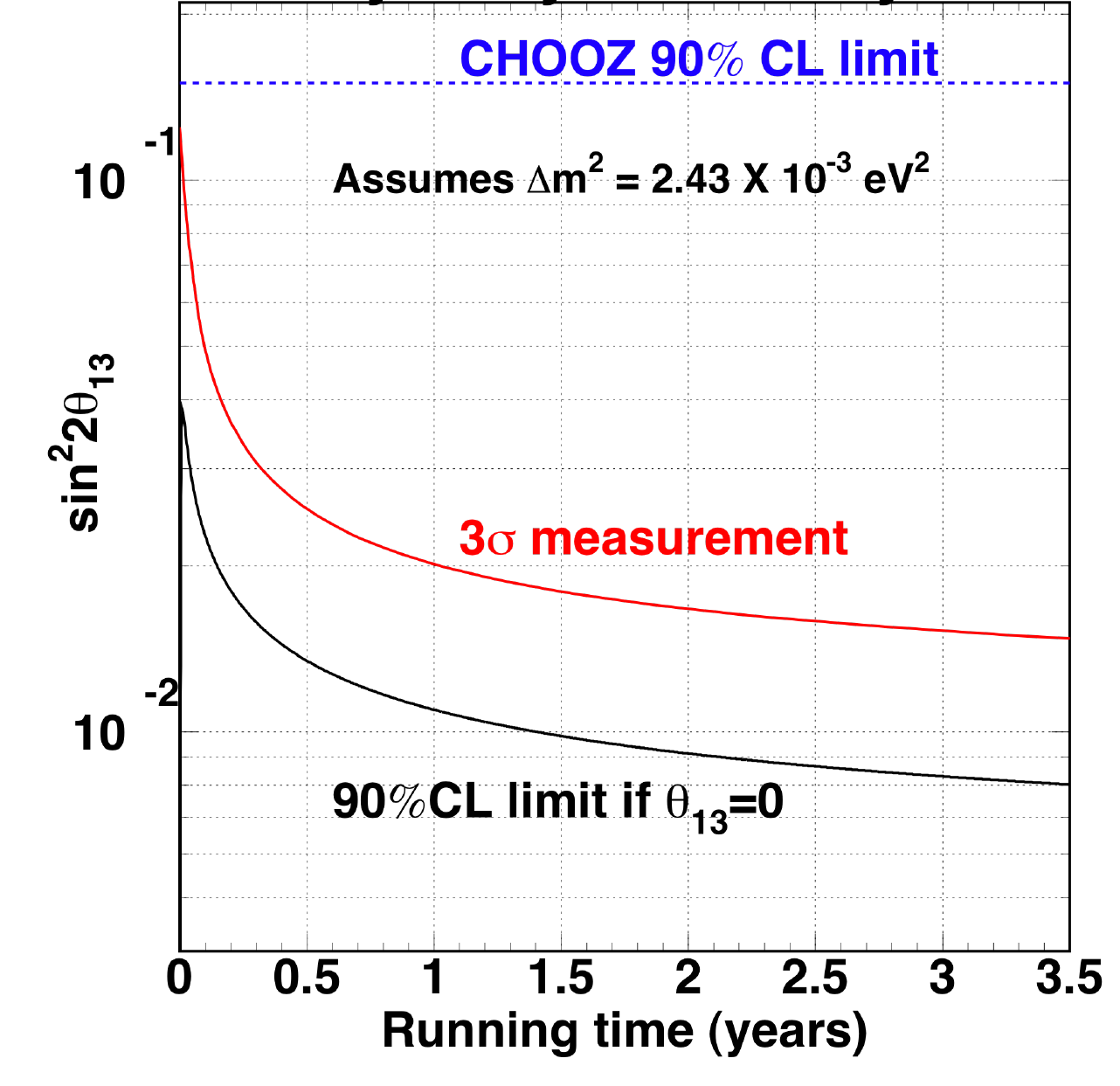}
\caption{Experiment sensitivity vs. time.} \label{fig4}
\end{minipage}
\end{figure}
%\vspace{-0.13in}

\section*{Acknowledgements}
\vspace{-0.05in}
This work was supported in part by the Ministry of Science and Technology of China
(contract no. 2006CB808100), the Chinese Academy of Sciences, the National Natural
Science Foundation of China (Project numbers 10890090 and 11005073), the Guangdong provincial
government, the Shenzhen Municipal government, the China Guangdong Nuclear Power
Group, the Ministry of Education of China, the Shanghai Jiao Tong University, the Research Grants Council of the Hong
Kong Special Administrative Region of China (Project numbers 400805, 703307, 704007 and
2300017), the focused investment scheme of CUHK and University Development Fund
of The University of Hong Kong, the MOE program for Research of Excellence at
National Taiwan University and NSC fund support, the United States Department of
Energy (DE-AC02-98CH10886, DE-AS02-98CH1-886, DE-FG02-92ER40709,
DE-FG02-07ER41518, DE-FG02-91ER40671, DE-FG02-08ER41575, DE-FG02-88ER40397, DE-FG02-95ER40896, DE-FG02-01ER41155 and DE-FG02-84ER40153), the U.S. National Science Foundation (Grants PHY-0653013,
PHY-0650979, PHY-0555674, PHY-0901954 and NSF03-54951), the Alfred P. Sloan Foundation, the
University of Wisconsin, the Virginia Polytechnic Institute and State University,
the Ministry of Education, Youth and Sports of the Czech Republic (Project numbers
MSM0021620859 and ME08076), the Czech Science Foundation (Project number
GACR202/08/0760),  and the Joint Institute of Nuclear Research in Dubna, Russia.

\end{document}